\documentclass[prb,twocolumn,aps,10pt,superscriptaddress]{revtex4-2}
\usepackage{amsmath}
\usepackage{amsfonts}
\usepackage{amssymb}
\usepackage{graphicx}
\usepackage{lmodern}
\usepackage{color}
\usepackage{mathtools}
\usepackage[pdfstartview=FitH,colorlinks=true,citecolor=blue,linkcolor=blue,urlcolor=blue]{hyperref}
\usepackage{float}

\begin{document}

\title{Waiting time fluctuations in quasi-one-dimensional disordered conductors}
\author{F.~Schulz}
\affiliation{Department of Physics, University of Basel, Klingelbergstrasse 82, CH-4056 Basel, Switzerland}
\author{M.~Albert}
\affiliation{Universit\'e C\^ote d'Azur, CNRS, Institut de Physique de Nice, 06200 Nice, France}
\affiliation{Institut Universitaire de France (IUF)}
\date{\today}
\begin{abstract}
We consider sample to sample fluctuations of the waiting time between the detection of two consecutive electrons in quasi-one-dimensional disordered conductors at zero temperature. We compute the full distribution of the mean waiting time along the crossover from ballistic to localised transport in the framework of the Dorokhov-Mello-Pereyra-Kumar theory for an arbitrary number of conduction channels. In particular we show that its variance, with respect to disorder, displays universal fluctuations similar to the universal conductance fluctuations in the metallic regime. We then discuss the statistical properties of the jitter associated to quantum fluctuations of the waiting time.
\end{abstract}

\maketitle

\section{Introduction}\label{sec.:introduction}

Electronic transport at the nanoscale is known to be stochastic due to the quantum nature of particles \cite{buttiker2000}. In the Landauer-B\"uttiker formalism, the complexity of a mesoscopic conductor can be encoded in a scattering matrix describing the different scattering processes from electronic modes of the connected leads \cite{buttiker1992}. While the average current is related to the sum of all the transmission probabilities, and already contains many non trivial information, granularity of charge carriers lead to a fundamental quantum noise called shot noise. At low frequency and temperature, the Full Counting Statistics (FCS) of transferred charges through the conductor, whose second moment is the shot noise, is usually a generalized binomial distribution \cite{Levitov1993}. At higher frequency, another source of noise has to be considered and is often referred as quantum jitter. It is related to the fact that even if an electron is transmitted through the sample, its time of detection is random due to its wave nature. This jitter contains essential information about the electronic quantum state and has important consequences in the field of electron quantum optics for instance \cite{Singleelectron2,Singleelectron3,Singleelectron4,Waintal2018}. In order to better characterize this fundamental noise, the Waiting Time Distribution (WTD), namely the probability distribution of time delays between the detection of two consecutive electrons, has been introduced \cite{Brandes2008,WTD1,WTD1b,WTD8,WTD11} and shown to be useful in many situations.   

If disorder is present in the sample, its interplay with wave coherence of the charge carriers may have drastic consequences. One extreme example is the absence of diffusion in low dimensional conductors due to destructive interference known as Anderson localization \cite{Anderson1958,Abrahams1979}, if the system size is larger than a typical localization length $\xi$. This is the localized regime where the conductance of the sample is exponentially suppressed and presents strong fluctuations. Another non trivial example is the existence of Universal Conductance Fluctuations (UCF) when the system size is much smaller than the localization length but much larger than the scattering mean free path $\ell$. In this metallic regime, the variance of the conductance has been predicted \cite{Lee1985} to be system independent and depends only of the existence or not of time reversal symmetry, which was experimentally observed \cite{Washburn1985,Mailly1992}. In general, the understanding of disordered conductors has been greatly improved by the application of random matrices to quantum transport \cite{Beenakker1997}. In particular, in the regime of quasi-one-dimensional transport, namely when transport is carried by a finite number of transverse modes, and weak disorder, the statistical distribution of the transmission coefficients has been shown to obey a universal scaling equation called the Dorokhov-Mello-Pereyra-Kumar (DMPK) equation \cite{DMPKoriginal}. This Fokker-Planck equation only depends on the number of channels $N$, the mean free path $\ell$ and the Dyson symmetry index $\beta$ which is equal to one for time reversal systems and two if time reversal symmetry is broken. It constitutes an universality class of disordered systems and has been successful to recover previous phenomena among others.

In this paper, we study the effect of disorder, in quasi-one-dimensional systems at zero temperature, on waiting times in the framework of the DMPK universality class. In Sec. \ref{sec.:model} we recall some important results on waiting times in clean conductors as well as general results for the statistical distribution of transmission coefficients in disordered systems. Then, Sec. \ref{sec.:results1} and \ref{sec.:results2} are devoted to the application of the DMPK theory to compute the probability density function, as well as sample to sample fluctuations of the mean waiting time. The latter being defined as the quantum average on each sample. In Sec. \ref{sec.:results3} we present results on the fluctuations of the temporal width of the WTD before giving our conclusions in Sec. \ref{sec.:conclusion}. Technical details are provided in the appendix.   

\section{Model}\label{sec.:model}

Before discussing our original results on the effect of disorder on waiting times, we make use of this section in order to properly define the quantities of interest, the model and recall some useful results for the reader. All along this work, we will consider one dimensional non interacting electrons at zero temperature in a two terminal geometry. The mesoscopic conductor will be modeled by a set of transmission coefficients $T_i$, $i=1,..,N$, $N$ being the number of channels. A small constant energy difference $eV$ is applied between the two leads (from left to right) so that transport is stationary and the energy dependence of the $T_i$ can be neglected. The constrains of zero temperature and energy dependence can potentially be lifted. For our purpose, we are concerned about left to right transmitted electrons. Since we are dealing with two types of averages, we will use the symbol $\langle\cdots \rangle$ to denote the average over quantum fluctuations and $\overline{\cdots}$ for the average over disorder. Similar, we use double brakets for the centred moments of waiting times and $\text{Var}(x)$ for the variance with respect to disorder. 

\subsection{Basic results on waiting times}\label{subsec.:modelWTD}

It is instructive to define the waiting times and their corresponding WTD in the simplest case of a single quantum channel \cite{WTD1b}. This would correspond, for instance, to the case of spin polarized electrons flowing through a quantum point contact of transmission $T$. In that case, the average current is simply $I=e^2 V T/h $ which can be interpreted as follow. On average, the left lead injects an electron every $\tau_V=h/eV$ due to the Pauli principle which is transmitted with probability $T$ \cite{Martin1992}. The noise is proportional to $T(1-T)$ and therefore vanishes for $T=1$ and the long time FCS is binomial with an attempt frequency $1/\tau_V$ and parameter $T$. The WTD is defined as the probability distribution of time delay $\tau$ between transmitted electrons. In that case, it has been shown to display a crossover from an exponential distribution close to pinch off ($T\ll 1$) to a Wigner surmise at perfect transmission \cite{WTD1b}. However, it always vanishes at $\tau=0$ which is the consequence of the Pauli principle. Due to the fermi statistics, the WTD also contains Friedel oscillations with period $\tau_V$ for $T<1$. 

When the number of conducting channels is increased, the vanishing probability at $\tau=0$ becomes finite because several non interacting electrons can be detected simultaneously. The random point process associated to the detection of transmitted electrons through the different channels is now a superposition of independent point processes similar to the single channel case but with a transmission $T_i$. In general, the average waiting time (with respect to quantum fluctuations) is given by
\begin{equation}\label{eq:meantau}
 \langle \tau \rangle=\frac{\tau_V}{\sum_i T_i} =\frac{\tau_V}{g},
\end{equation}
with $g$ being the dimensionless conductance.
It was shown that for a large number of channels, the WTD becomes exponential \cite{WTD2}.
 In the rest of the paper we will use $\tau_V$ as unit of time and drop it from now on. Hence, the mean waiting time is directly connected to the inverse of the dimensionless conductance, namely the dimensionless resistance $\rho=1/g$.

We now discuss typical deviations of the waiting time $\tau$, namely its variance with respect to quantum fluctuations. Although no analytical formula is known, we show in App. \ref{sec.:App1} that the following formula is very precise

\begin{equation}\label{eq:meantau2}
\langle\langle \tau^2\rangle\rangle=\langle\tau^2\rangle-\langle\tau\rangle^2=\frac{1}{(\sum_i T_i)^2}+\frac{a_N}{\sum_i T_i}  
\end{equation}
with $a_N$ being a channel dependent constant that decreases faster then $N$, such that the second term vanishes in the limit of $N\gg 1$. The square root of it, namely the standard deviation of $\tau$ with respect to quantum fluctuation, is a relevant measure of the quantum jitter.

\subsection{DMPK equation}\label{subsec.:modelDMPK}

We start this section by recalling some important results about the statistical distribution of transmission coefficients in a quasi-one-dimensional disordered conductors in the framework of the DMPK theory \cite{DMPKoriginal,Beenakker1997}. This theory describes the evolution of the joint probability of $T_i$ with respect to the longitudinal system size $L$, valid for weak disorder. 

It is customary to parameterize the transmission coefficients as $T_i=1/(1+\lambda_i)$ or $T_i=1/\cosh^2 x_i$ with $\lambda_i$ or $x_i$ positive real numbers. The DMPK equation in terms of $\lambda_i$ reads

\begin{eqnarray}
  \ell \frac{\partial P}{\partial L}=\frac{2}{\gamma}\sum_{i=1}^N \frac{\partial}{\partial \lambda_i} \lambda_i(1+\lambda_i)J\frac{\partial}{\partial\lambda_i}\frac{P}{J}\\
  J=\prod_{i=1}^N \prod_{j=i+1}^N |\lambda_j-\lambda_i|^\beta,
\end{eqnarray}
with $\gamma=\beta(N-1)+2$ and $\beta$ being the universal symmetry index denoting the presence ($\beta=1$) and absence ($\beta=2$) of time reversal symmetry. The only microscopic parameter entering this equation is $\ell$ the mean free path. This equation has to be complemented with the boundary condition $P(\{\lambda_i\},L\to 0)\to \prod_i \delta(\lambda_i-0^+)$ which describes the ballistic limit. It is important to note that although the electron transport occurs through independent channels, the corresponding transmission coefficients are strongly correlated. This can be easily understood from the presence of the Jacobian $J$ which describes universal repulsion between the $\lambda_i$. 

Generically, the different regimes of transport are labeled as the metallic regime if $L\ll \gamma\ell$, or the localized regime if $L>\gamma\ell$. In the former case there is an extra distinction if $L<\ell$, namely the ballistic regime or if $\ell\ll L\ll \gamma\ell$ the metallic diffusive regime.

While Beenakker and Rejaei \cite{Beenakker1993} presented an exact solution for the situation with broken time reversal symmetry,  Caselle \cite{Caselle1994} extended the solution for arbitrary $\beta$. It was further shown that the solution of the DMPK-equation in the metallic regime has the form of a Gibbs distribution, $P(\{x_i\})\propto \exp[-\beta \mathcal{H}(\{x_i\})$, with
\begin{eqnarray}\label{eq.:16}
\mathcal{H}(\{x_i\})&&=\sum_{i<j}U(x_i,x_j)+\sum_iV(x_i),\\
U(x_i,x_j)&&=-\frac{1}{2}\left(\ln|\sinh^2x_j-\sinh^2x_i|+\ln|x_j^2-x_i^2|\right),\nonumber\\
V(x_i)&&=\frac{\gamma}{2\beta s}x_i^2-\frac{1}{2\beta}\ln|x_i\sinh 2x_i|,\nonumber
\end{eqnarray}
with $s=L/\ell$. This approximation turned out to be very precise within every regime and was used to compute the conductance distribution numerically using Monte-Carlo sampling \cite{Mello2002}. The Hamiltonian $\mathcal{H}$ in Eq.~(\ref{eq.:16}) can be viewed as the one of $N$ interacting classical particles located at position $x_i$, within a confinement potential $V(x_i)$ and the particle interaction $U(x_i,x_j)$. The relation between the positions and the transmission values being $T_i=1/\cosh^2(x_i)$. This is the method we employ in this paper to compute the average over disorder, while we usually average over $10^7$ configurations.

In addition, there are some limiting cases where the distribution can be calculated analytically. The first one is the single channel limit $N=1$ where an exact integral solution has been obtained by Gertsenshtein and Vasil'ev \cite{Vasilev1959}, or Abrikosov \cite{Abrikosov1981}. In terms of the inverse conductance $\rho=\langle \tau \rangle=1/g$ the solution reads

\begin{equation}\label{eq:abrikosov}
  P(\rho,s)=\frac{2}{\sqrt{\pi s^3}}\int_{\mathrm{arch}\sqrt{\rho}}^{+\infty} \frac{x \exp[-(x^2/s+s/4)]}{(\mathrm{ch}^2 x-\rho)^{1/2}}\,dx,
\end{equation}
resulting in an exponentially increasing $\overline{\rho}(s)$.
Further, in the localized regime, for arbitrary $N$, the Coulomb gas becomes very dilute but crystallized, such that the $x_i$ are very large and strongly separated \cite{Beenakker1997}. The joint probability of the $x_i$ factorizes to a product of shifted Gaussian distributions
\begin{equation}\label{eq:localized}
  P(\{x_i\})\simeq \left(\frac{\gamma \ell}{2 \pi L}\right)^{N/2}\prod_{i=1}^N \exp\left[-\frac{\gamma\ell}{2L}(x_i-L/\xi_i)^2\right]
\end{equation}
with $\xi_i=\gamma\ell/(1+\beta i-\beta)$. Deeply in the localized regime, the conductance is dominated by the smallest $x_i$ and is distributed according to a log normal distribution.

Most strikingly, in the diffusive metallic regime, the conductance distribution is approximately Gaussian with $\overline g=N/s+\frac{1}{3}(1-2/\beta)$ and UCF $\mathrm{Var}(g)=\frac{2}{15\beta}$.  

\section{Full distribution of the Mean waiting time}\label{sec.:results1}
\begin{figure}
\begin{center}
\includegraphics[width=1.0\linewidth]{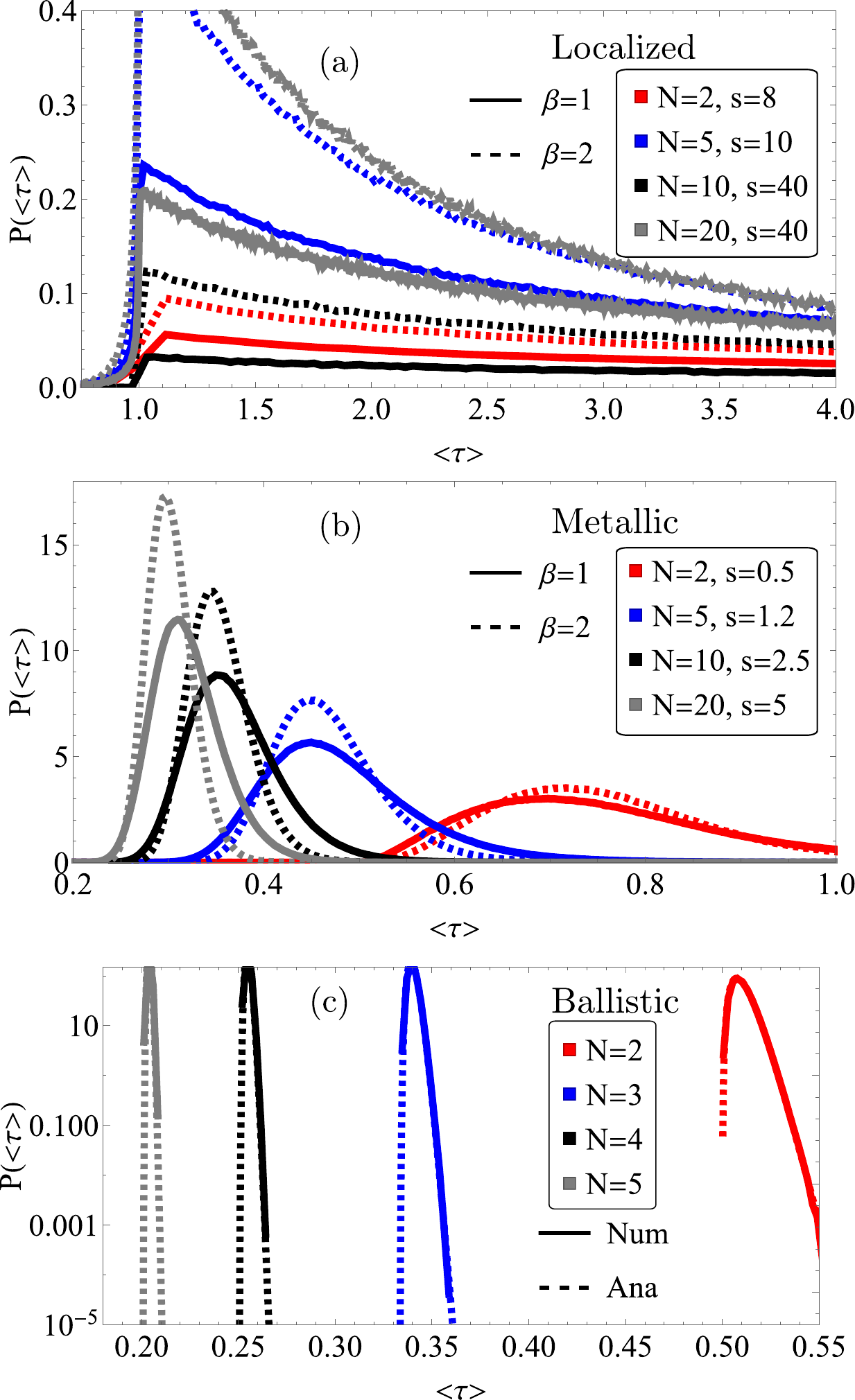}
\caption{(Color online) Probability density function of the mean waiting time $\langle\tau\rangle$ for several transmission channels. $(a) [(b)]$ presents the localized [metallic] regime, for several combinations of $s$ and $N$ with $\beta=1$ and $2$, respectively. $(c)$ shows the ballistic regime for $s=0.02$ and $\beta=2$ for the numerical Monte-Carlo simulations (solid lines), as well as the analytical approach (dashed lines) from Eq.~(\ref{eq.:ptaubal}).}
\label{fig.:ptau}
\end{center}
\end{figure}

We present in this section results obtained from our comprehensive study to provide valuable insights into the probability density function of the mean waiting time $\langle\tau\rangle$ contributing to the understanding of the regimes of a ballistic, metallic and localized one dimensional conductors. The general approach for the probability density function can be calculated by 
\begin{equation}
P( \langle\tau\rangle )=\left[\prod _i \int_0^\infty dx_i\right]\delta\left(\langle\tau\rangle-\frac{1}{\sum_i 1/\cosh [x_i]^2 }\right)P(\{x_i\}),
\end{equation} 
where $\delta(x)$ is the Dirac-delta distribution. We begin with the Monte-Carlo sampling within the localized regime, where $s/\gamma \gg1$ and all transmission values $x_1\ll x_2\ll x_3...\ll x_n$ are strongly separated. 
The distributions of the mean waiting time $P( \langle\tau\rangle )$ in the localized regime for $\beta=1$ and 2 are shown in Fig.~\ref{fig.:ptau}(a). 
Deep in the localized regime we find that independent of the number of transmitting channels, the tail of the distributions follows a Log-Normal distribution with a lower bound cut off at $\langle\tau\rangle=1$, given by 
\begin{equation}\label{eq.:ptauloc}
  p(\langle \tau \rangle)\simeq \frac{1}{2\langle\tau\rangle}\left(\frac{\gamma}{2\pi s}\right)^{1/2}\exp\left[-\frac{\gamma}{2s}(\ln(4\langle \tau\rangle)/2-s/\gamma)^2\right].
\end{equation}
Such a cut off has also be found for the probability density function of the conductance \cite{woelfle1999}.
Thus, in the strongly localized regime the localization length is much smaller then the sample length and all transmitting channels are strongly suppressed. In this limit the transport is dominated by a single channel [see Eq.~\eqref{eq.:ptauloc}], namely the channel with the smallest $x_i$. Conclusively, the localized single channel calculations by Abrikosov \cite{Abrikosov1981,Beenakker1997} become the relevant description.

By downsizing the length of the sample we change the setup into the metallic regime (for reasonable $N$). The results in that regime are shown in Fig.~\ref{fig.:ptau}(b). There, the probability density functions for $N=10$ and $N=20$ represent a Gaussian, distributed around $\overline{\langle\tau\rangle}$. Similar results have been found for conductance distribution \cite{Beenakker1997,Muttalib2003}. For $N=2$ (red lines), $P(\langle\tau\rangle)$, with its maximum at $\overline{\langle\tau\rangle}$, shows the direct conversion of the probability density function from the localized and ballistic regime, since the metallic regime is impassable. Similar, for $N=5$ the results are not Gaussian, but represent the crossover into the metallic situation. In general, the metallic regime is reached more favourable for $\beta=2$, since the localization length $\gamma\ell$ becomes larger in TRS broken systems. 

Next, we reduce the system length to the ballistic regime ($s\ll 1$), where each channel becomes almost fully conducting. We highlight the corresponding distributions in Fig.~\ref{fig.:ptau}(c).
For clarity we show the distributions on a Log-scale for $\beta=2$. There, the solid lines represent the analytical results that we derive in App.~\ref{sec.:App2}, given by
\begin{eqnarray}\label{eq.:ptaubal}
P(\langle \tau \rangle)=\frac{\delta ^{\alpha } \left(\langle \tau \rangle -\frac{1}{N}\right)^{\alpha -1} e^{-\delta(\langle \tau \rangle -\frac{1}{N})}}{\Gamma (\alpha )},
\end{eqnarray}
which is a Gamma distribution with the shape parameter $\alpha=\frac{\gamma N}{2}$, the rate parameter $\delta=N\alpha/s$ and $\Gamma (x)$ the Gamma function. In the ballistic situation we find excellent agreement with the analytical formula for $s\ll1$, also for $\beta=1$ and any $N$ (not shown). Physically, the transmission channels become independent of each other, due to the fact that the system length is smaller then the mean free path. The limit of Eq.~(\ref{eq.:ptaubal}) to $N=1$ is also in very good agreement with the solution made by Abrikosov [see Eq.~(\ref{eq:abrikosov})].
As a side remark, the probability density function for the transmission values in the ballistic regime is for $x$ represented by the positive eigenvalues of a random chiral matrix ensemble for the symmetry classes $\beta$ \cite{ForresterRMT}. 
For details we refer to App.~\ref{sec.:App2}.

\section{Universal waiting time fluctuations}\label{sec.:results2}

\begin{figure}
\begin{center}
\includegraphics[width=0.90\linewidth]{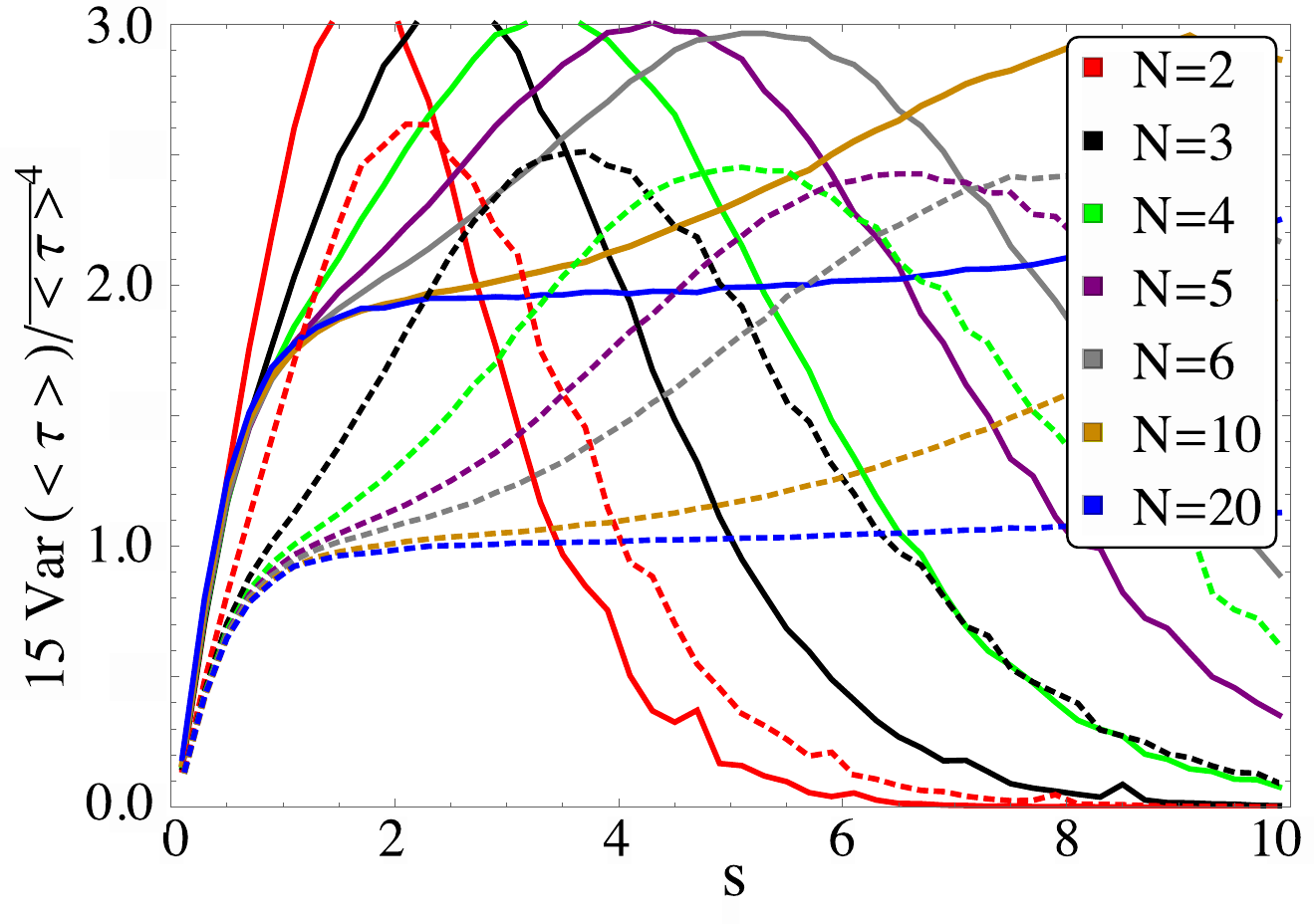}
\caption{(Color online) Rescaled variance of the mean waiting time $\langle\tau\rangle$ for several transmission channels for increasing length of the setup. The symmetry classes $\beta=1 (2)$ are represented by solid (dashed) lines.}
\label{fig.:varTau}
\end{center}
\end{figure}

\begin{figure}
\begin{center}
\includegraphics[width=1.\linewidth]{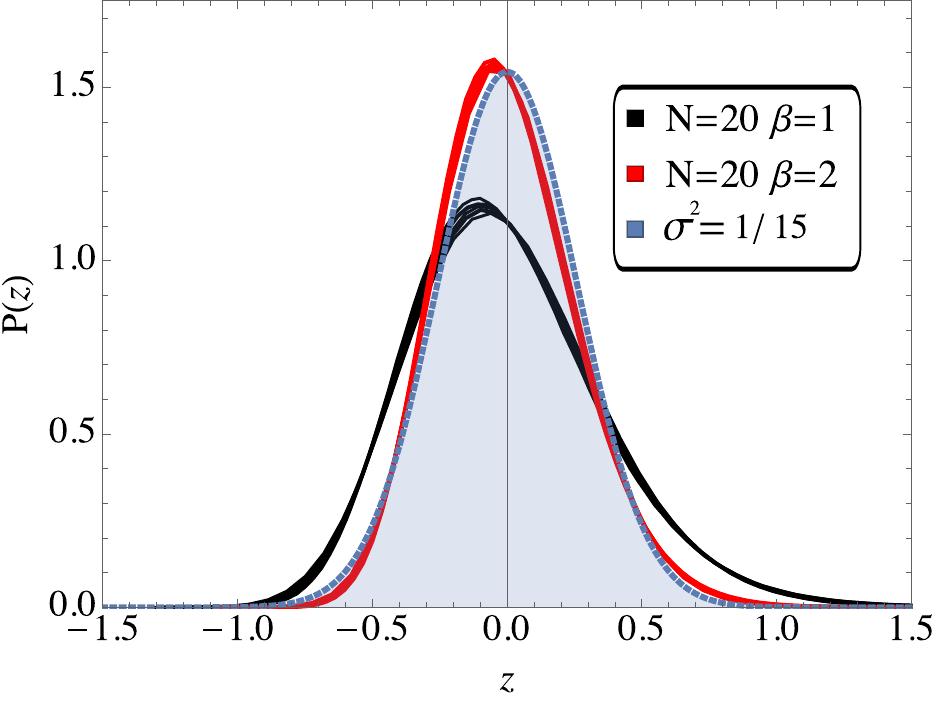}
\caption{(Color online) Probability density of the scaled mean waiting time $z=(\langle\tau\rangle- \overline{\langle\tau\rangle})/\overline{\langle\tau\rangle}^2$ within the metallic regime. The length of the setup is varied in the range of $s \in(2.9,4.9)$, for $N=20$. The symmetries for $\beta=1(2)$ are shown in black (red), while the blue curve represents the expected normal distribution for $\beta=2$.}
\label{fig.:Pz-Gauss}
\end{center}
\end{figure}

In this section, we present distinct sample to sample fluctuations of the mean waiting across various configurations of the sample, revealing the connection to the UCF \cite{Lee1985,Beenakker1997,Mello2002}. By calculating the expectation values for $\overline{\langle\tau\rangle}$ and $\overline{\langle\tau\rangle^2}$, the variance of the mean waiting time can be performed by 
$\text{Var}(\langle\tau\rangle)=\overline{\langle\tau\rangle^2}-\overline{\langle\tau\rangle}^2$.
We evaluate the corresponding variance in dependence of $s$, shown in Fig.~\ref{fig.:varTau}.
There we find, independent of the number of channels a vanishing variance around $s\approx 0$, corresponding to perfect transmission of non fluctuating transmission coefficients. Similarly, at very large $s$, the transmission coefficients become independent from each other, resulting in an effective single channel with $\text{Var}(\langle\tau\rangle)<\overline{\langle\tau\rangle^4}$. Most interesting, at length scales, where the correlations between the channels are strong, an increase of the number of channels results from a single maxima (see red curves for $N=2$) to an almost constant value of the variance (see brown or blue curves for $N=10$ or $N=20$, respectively). This fascinating result has previously shown in the UCF \cite{Lee1985,Beenakker1997,Mello2002}. It is well known that in the diffusive metallic regime the transmission density follows a bimodal distribution [$P(T)\propto 1/(T\sqrt{1-T})$], such that each of the channels has the probability of either perfect transmission or zero transmission \cite{Beenakker1992}.  
The correspondence to the disorder averaged mean waiting time is found by the following argument. In the metallic regime we assume that $\overline{\langle\tau\rangle}=\frac{1}{\overline{g}+\delta g}$, where $\delta g$ are small fluctuations around $\overline{g}$, leading to $\overline{\langle\tau\rangle}\approx (1-\frac{\delta g}{\overline{g}})/\overline{g}$, such that $\text{Var} (\langle\tau\rangle)\approx\frac{\text{Var}(g)}{\overline{g}^4}=\frac{2}{15\beta}\overline{\langle\tau\rangle}^4$. Conclusively, the variance of the mean waiting time contains the well known symmetry dependent constant $\frac{2}{15\beta}$, while on the same time it depends on the fourth power of the disorder averaged mean waiting time. This feature is a consequence of the Gaussian distribution of $\langle\tau\rangle$ within the metallic regime. We emphasise this by presenting the transformed PDF $P(z)$, where $z=(\langle\tau\rangle- \overline{\langle\tau\rangle})/\overline{\langle\tau\rangle}^2$, for 20 channels in Fig. \ref{fig.:Pz-Gauss}.
There, we find in the metallic regime ($s$ is varied from $2.9$ to $4.9$) an approximate Gaussian distribution, with variance $\sigma^2=\frac{2}{15\beta}$. The red curves in Fig. \ref{fig.:Pz-Gauss} corresponds to the TRS broken situation ($\beta=2$), which is in very good agreement with the analytics (blue curve). For $\beta=1$ one already finds deviations from that, due to the fact that the metallic regime is not perfectly reached. The results can be improved with an increasing number of channels ($N>20$). 
In summary, the waiting time fluctuations are reminiscent of UCF \cite{Mello2002}, but cannot be considered universal due to the rescaling with $\overline{\langle\tau\rangle}^4$. 

\section{Quantum jitter}\label{sec.:results3}

In the previous sections we have discussed the statistical properties of the mean waiting time $\langle\tau\rangle$ with respect to disorder. It is however natural to wonder about the ones of higher order moments of the WTD $\langle \tau^k\rangle$ with $k$ an integer number. In this work we focus on the second moment which is the first characterization of quantum fluctuations of the waiting times. As discussed in Sec. \ref{sec.:model} and App. \ref{sec.:App1} it can be written analytically according to Eq. (\ref{eq:meantau2}) and therefore depends on the statistical properties of the inverse of the dimensionless conductance and its square. This brings information which are beyond the scope of linear statistics \cite{Beenakker1997}.

\begin{figure}
\begin{center}
\includegraphics[width=0.85\linewidth]{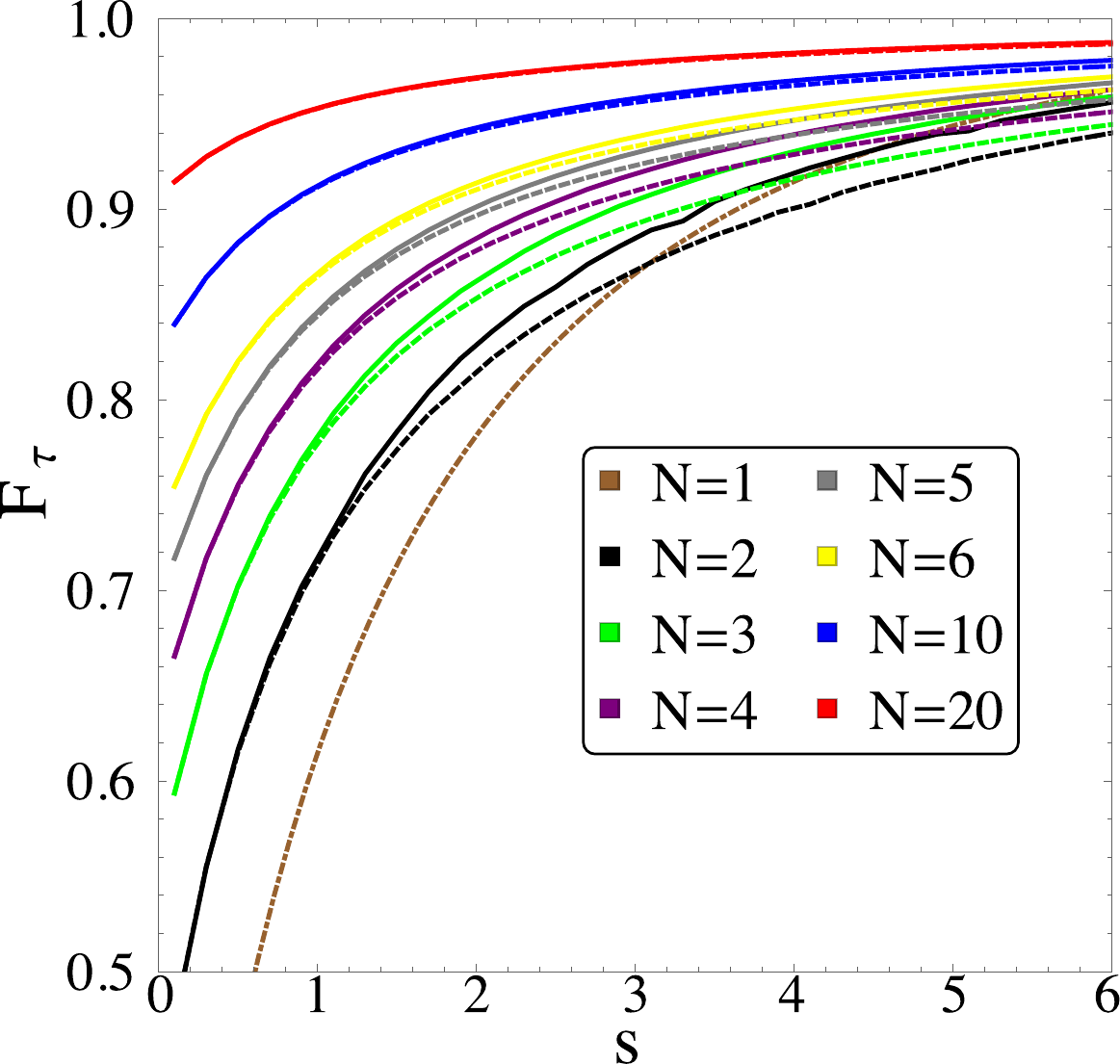}
\caption{(Color online) Fano factor $F_\tau=\overline{\langle\tau^2\rangle/\langle\tau\rangle^2}-1$ for several transmission channels for increasing length of the setup. The symmetry classes $\beta=1 (2)$ are represented by solid (dashed) lines. The brown dotted dashed line corresponds to $N=1$ (independent of $\beta$).}
\label{fig.:FanoS}
\end{center}
\end{figure}

In Fig. \ref{fig.:FanoS} we present results about $\overline{\langle\tau^2\rangle}-\overline{\langle\tau\rangle^2}$, namely the average over disorder of the second cumulant of the WTD. More precisely we compute 

\begin{equation}\label{F_tau}
    F_\tau=\overline{\langle\tau^2\rangle/\langle\tau\rangle^2}-1
\end{equation}
 in the spirit of Ref. \cite{WTD1b} in order to compare it to the Fano factor which is the ratio of the shot noise and the average current. Note that we could also look at $\overline{\langle\tau^2\rangle}/\overline{\langle\tau\rangle^2}$ which displays similar features. The two Fano factors are known to match for renewal processes, namely when consecutive waiting times are uncorrelated. This regime is known to be attained only in the limit of small transparency  \cite{WTD1b} and therefore this connection will be valid in the localized regime only. In order to understand this it is instructive to consider the single channel case without disorder. In that case, the Fano factor of the FCS is simply $F=T(1-T)/T=1-T$ and therefore is one in the tunneling limit where FCS is poissonian and cancel for perfect transmission. However, the second moment of the WTD is finite in that case. This missmatch between the two quantities is a signature of the quantum jitter of electrons.

We now go back to the discussion of quasi-one-dimensional disordered systems. In Fig.~\ref{fig.:FanoS}, in the localized regime ($s\gg 1$), we indeed recover that $F_\tau$ tends to one and therefore match the results of FCS. This is expected since transport of electrons becomes poissonian.

As the system enters the metallic regime for smaller values of $s$ and reasonable large $N$, FCS would predict a universal value of $1/3$ \cite{Beenakker1992,Yakovets1995}. This is not what we observe for $F_\tau$ in accordance with the prediction of Eq. (\ref{fig.:varTau}). This equation predicts that $F_\tau=1-a_N \overline{g}$. We show in Appendix \ref{sec.:App1} that $a_N$ decays faster than $N$ while $\overline{g}$ scales as $N$ in the metallic diffusive regime. For large $N$, $F_\tau$ then tends to one which is larger than the $1/3$ prediction. This result is surprising at first sight since electron transport is described by the superposition of a large number of point processes. The consecutive waiting times are then expected to be uncorrelated. However, this leads to an exponential WTD, which cannot reproduce the bimodal statistics of the FCS. In that regime, the long range rigidity of the electronic stream in each channel cannot be described by the WTD alone and the renewal assumption does not hold at all. The WTD and the FCS are therefore sensitive to different aspects of quantum transport.

Finally, as the systems approaches the ballistic regime, $F_\tau$ decreases but never vanishes as it should for the FCS Fano factor. This is the regime where the quantum jitter is dominant and is highligthed by waiting time fluctuations \cite{WTD1b}. At exactly $s=0$ the system is perfectly ballistic and can be interpret as the clean limit of perfectly transmitting channels ($F_\tau(s=0)=1-a_nN$). Moreover it is interesting to note that for $s<1$ the Fano factor becomes $\beta$ independent which is an expected results since TRS is no longer important when the number of scattering events is of order one. This is a regime where the FCS cannot bring any interesting information while the WTD reveals interesting features encoded in the electronic many body state.

\section{Conclusion}\label{sec.:conclusion}
In summary, we discussed in this work the sample to sample fluctuations of the waiting times due to disorder for quasi-one-dimensional mesoscopic conductors in the framework of the DMPK equation. 

We have computed numerically the full distribution of the mean waiting time $\langle\tau\rangle$ along the cross-over from the ballistic regime to the localized regime and have given analytical expressions in all limiting cases. In addition, we have found that the variance of the mean waiting time, in the diffusive metallic regime, displays fluctuations reminiscent to the universal conductance fluctuations.

We have also studied the statistical properties of the quantum fluctuations of the waiting times. We further showed and discussed important discrepancies between the statistics of the waiting time and full counting statistics.

A natural extension of this work would be to look at the statistics of higher order cumulants of the waiting times. In the spirit of this study it would also be interesting to look at the waiting time fluctuations in chaotic quantum dots where the distribution of transmission coefficient is given by RMT \cite{Beenakker1997}.

\section*{Acknowledgments}
We would like to thank F. H\'ebert for helpful discussions about classical Monte-Carlo simulations and Aur\'elien Grabsch about random matrix theory. This work has been supported by the French government, through the UCA$^\mathrm{JEDI}$ Investments in the Future project managed by the National Research Agency (ANR) with the reference number ANR-15-IDEX-01.
\appendix

\section{Analytical expression for the jitter}\label{sec.:App1}

We present in this appendix an analytical expression for the second moment of the WTD which reproduces with great accuracy the exact numerical results. Before doing this we recap the method to compute the WTD.

\begin{figure}
\begin{center}
\includegraphics[width=0.9\linewidth]{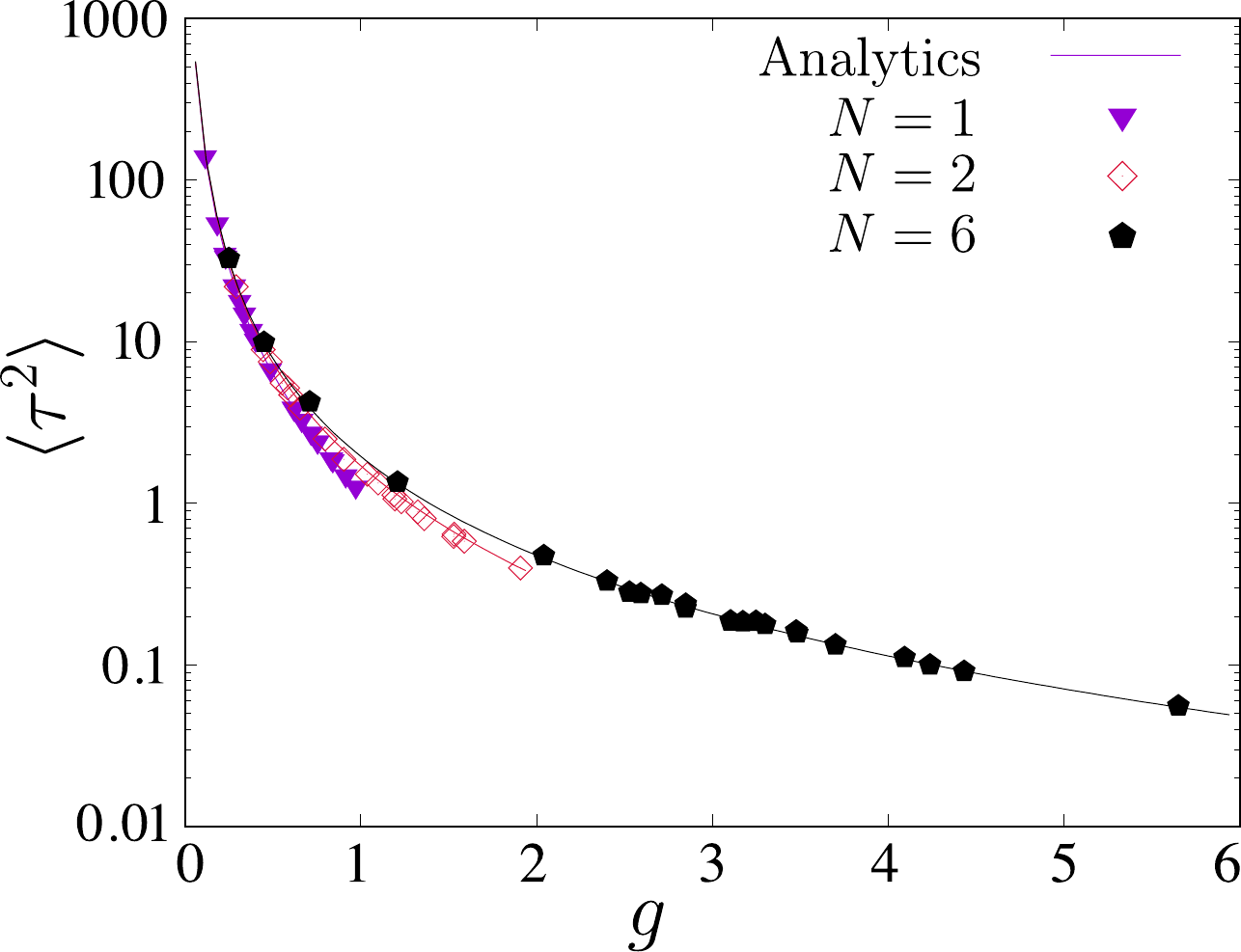}
\caption{Second moment $\langle \tau^2 \rangle$ of the WTD as a function of the conductance $g$ for different number of channels $N$. The symbols correspond to numerical evaluation of Eq. (\ref{WTDeq.:tau2num}) while the full lines are given by Eq. (\ref{WTDeq.:14}).}
\label{fig.:app1}
\end{center}
\end{figure}

The WTD $\mathcal W(\tau)$ denotes the probability distribution for the time delay $\tau$ between the detection of two consecutive electrons. In general, it is customary to compute the WTD from the idle time probability $\Pi(\tau)$, namely the probability of not detecting any electron during a period of measurement $\tau$. For a single channel and for non-interacting electrons at zero temperature $\Pi(\tau)$ can be expressed as a determinant which depends on the transmission probability $T$ of the channel \cite{WTD1b}. In the limit of perfect transmission or close to pinch off ($T\ll 1$) the latter can be approximated by (in units of $\tau_V=h/eV)$
\begin{eqnarray}
\Pi_\text{W}(\tau)&&= e^{-\frac{4 \tau ^2}{\pi  \bar{\tau}^2}}-\tau  \text{erfc}\left(\frac{2 \tau }{\sqrt{\pi }}\right)\\
\Pi_\text{P}(\tau)&&=e^{-\tau}.
\end{eqnarray}
The WTD distribution is obtained through
\begin{equation}
    \mathcal{W}(\tau)=\langle\tau\rangle\frac{d^2 \Pi(\tau)}{d\tau^2},
\end{equation}
with $\langle\tau\rangle=-1/\Pi'(0)$.
When transport occurs through $N$ channels with transmission probability $T_i$, the idle time probability factorizes and reads 
\begin{equation}\label{WTDeq.:12}
\Pi_N(\tau)=\prod_{i=1}^N \Pi(\tau,T_i),
\end{equation} 
where each $\Pi(\tau,T_i)$ is computed from the determinant formula given in Ref. \cite{WTD1b}. The first moment is shown to be $\langle\tau\rangle=1/\sum_i T_i$ which is Eq. (\ref{eq:meantau}) of the main text \cite{WTD2}. The second moment $\langle\tau^2\rangle$, for $N$ channels, can be calculated (after integration by parts) as
\begin{equation}\label{WTDeq.:tau2num}
  \langle\tau^2\rangle=2\langle\tau\rangle \int_0^\infty \Pi_N(\tau)d\tau 
\end{equation}
It is obvious that the second moment must be a symmetric function of the $T_i$. We therefore expand it as a power series of $1/g$, with $g=\sum_i T_i$. In order to recover the limiting cases it has to stop to second order in $1/g$. We therefore take the following ansatz for $\langle\tau^2\rangle$

 \begin{equation}
 \label{WTDeq.:14}
\langle\tau^2 \rangle = \frac{a_N}{\sum_i T_i}+\frac{b_N}{(\sum_i T_i)^2},\\
\end{equation}  
where $a_N$ and $b_N$ are constants depending on the number of channels. The constant $a_N$ and $b_N$ are found by matching this result with the limiting cases $T_i=1$ and $T_i\ll 1$. We find $b_N=2$ and $a_N=2\int_0^\infty d\tau \Pi_\text{W}(\tau)-2/N$. This gives for instance $a_1=3\pi/8-2$, $a_2=(19\sqrt{2}-16)\pi/48-1$. For larger values of $N$, $a_N\simeq \sqrt{2\pi/N}e^{N/2}\text{erfc}(\sqrt{N/2})-2/N$ which behaves as $-2/N^2$ for $N\gg 1$. In Fig. \ref{fig.:app1} we show some examples of comparison between the analytical ansatz and the numerical evaluation of $\langle\tau^2\rangle$ which demonstrate a very good agreement. We have tested this formula up to $N=20$ channels (not shown) with the same conclusion.

\section{Probability density function in the ballistic regime}\label{sec.:App2}
We present here the calculation for the distribution of the mean waiting time $\langle \tau \rangle$ within the ballistic regime $s\ll1$. We start with Eq.~\eqref{eq.:16} and assume that each transmission value is close to unity, such that we can approximate
\begin{eqnarray}
P(\{x_i\})\propto \prod_{i<j}|x_j^2-x_i^2|^\beta\prod_i\sqrt{2x_i^2}e^{-\frac{\gamma}{2s}x_i^2},
\end{eqnarray}
as well as 
\begin{eqnarray}
     \frac{1}{g}=\frac{1}{\sum_i 1/\cosh{x_i}^2}\approx\frac{1}{N}\left(1+\frac{1}{N}\sum_i x_i^2\right).
\end{eqnarray}
The distribution for $\langle \tau \rangle$ can be calculated by
\begin{eqnarray}
    P(\langle \tau \rangle)=\prod_i\int dx_i\delta \left[\langle \tau \rangle-\frac{1}{N}-\sum_i \frac{x_i^2}{N^2}\right]P(\{x_i\}),
\end{eqnarray}
which can be carried out within spherical coordinates. The integration over the angles results in a constant term that we include within the normalization condition, such that we are left ($\langle \tau \rangle\geq\frac{1}{N}$) with 
\begin{eqnarray}
 P(\langle \tau \rangle)\propto \int dr   \delta \left(\langle \tau \rangle-\frac{1}{N}-\frac{r^2}{N^2}\right) r^{N\gamma-1} e^{-\frac{\gamma  r^2}{2 s}}\nonumber \\
 =\frac{1}{2} N^\alpha  ( \langle \tau \rangle-\frac{1}{N})^{\alpha-1} e^{-\frac{\alpha  N ( \langle \tau \rangle-\frac{1}{N})}{ s}},
\end{eqnarray}
with $\alpha=\frac{\gamma N}{2}$.
After normalization the PDF is in compact form given by 
\begin{eqnarray}
P(\langle \tau \rangle)=\frac{\delta ^{\alpha } \left(\langle \tau \rangle -\frac{1}{N}\right)^{\alpha -1} e^{-\delta(\langle \tau \rangle -\frac{1}{N})}}{\Gamma (\alpha )},
\end{eqnarray}
which is the Gamma distribution with the shape parameter $\alpha$ and the rate $\delta=N\alpha/s$.


\end{document}